\documentstyle{aipproc}
\begin{document} 
\title{UHECR Production and Curvature TeV Emission
in Nearby, Dormant AGNs}
\author{Amir Levinson}
\address{School of Physics and Astronomy, Tel Aviv University,
Tel Aviv 69978, Israel} 
\maketitle
 
\begin{abstract}

The possibility that rapidly rotating supermassive black holes associated
with quasar remnants may provide sites for the acceleration of 
ultra-high energy cosmic rays (UHECRs) is considered.
It is shown that prodigious TeV emission through curvature losses is
an important consequence of this mechanism.  Given the measured
UHECR flux, it is expected that nearby dormant AGNs will be detectable 
by current Tev experiments.  The 
implications for the Sgr A* source are also briefly considered.
\end{abstract}

\section{Introduction}

Large efforts have been devoted in recent years to explore the origin of
the ultra-high energy cosmic rays (UHECRs).  Two distinctly different scenarios
are commonly considered; the Top-Down and Bottom-Up scenarios.  In the former one, 
the origin of the UHECRs is associated with the decay of some supermassive X 
particles (for a review see e.g., ref. \cite{BS99}), whereas in the latter scenario 
the UHECRs are assumed to be accelerated by astrophysical objects.  The main 
challenge confronted by the Bottom-Up scenario is to explain the apparent lack 
of a GZK cutoff \cite{G,ZK} in the present data, and to identify counterparts 
that coincide with the arrival directions of observed UHECRs.  Several classes of 
UHECR sources have been proposed, including: AGNs, GRBs, and neutron stars (see ref.
\cite{Ol00} and references therein).  
However, at present there seems to be no clear association of UHECR events with 
any of these objects.
In the following we consider further a recent idea, put forward by Boldt and 
Ghosh \cite{BG99}, that the UHECRs are accelerated by the supermassive black 
holes that appear to be present in the centers of normal galaxies \cite{M98}.
Such systems may represent quasar remnants or dormant AGNs that were
perhaps active during earlier phases in their evolution.

\section{Particle Acceleration}
A rotating black hole of 
mass $M=10^9M_9 M_{\sun}$ and specific angular momentum $a$ (measured in 
units of $c^2/G$), threaded by magnetic 
field of strength $B=10^4B_4$ Gauss, can induce an {\it emf} of
\begin{equation}
\Delta V\sim 4\times10^{20} (a/M)B_4M_9(h/R_g)^2\ \ \ \ {\rm volts},
\label{V}
\end{equation} 
where $h$ is the gap height, and $R_g=GM/c^2$ is the gravitational radius.
As pointed out in ref. \cite{BG99}, for the parameters corresponding to 
galaxies having $M_9>1$ in the list presented in ref. \cite{M98}, the 
electric potential given by the above equation is more than sufficient 
to accelerate the UHECRs observed at Earth, provided
that the {\it emf} is not shorted out. 

Now, in the case of AGNs, vacuum breakdown is likely to occur 
through the generation of pair cascades by the agency of the accretion
disk radiation, ultimately leading to the formation of 
magnetically dominated outflows via the BZ process \cite{BZ77} with 
a maximum power on the order of $10^{46}(a/M)^2M_9^2B_4^2$ erg s$^{-1}$ 
\cite{BZ77,TPM}.  These outflows are commonly associated with the powerful
jets often seen in blazars.  However, as shown in ref. \cite{Le00}, vacuum 
breakdown is not expected to occur in dormant AGNs (assuming that 
the electric field is not screened out by some surrounding plasma).  This 
then suggests that those systems, instead of producing luminous
radio jets as seen in blazars, may serve as accelerators of a 
small number of particles to ultra-high energies \cite{BG99}, provided
that the associated black holes were spun up to nearly their maximal 
spins during a phase when the dormant AGNs were active.

The UHECR production efficiency implied by the observed CR spectrum above 
the ankle is rather small; the corresponding UHECR power is given
approximately by  
\begin{equation}
L_{UHECR}=2\times 10^{42}\left(\frac{n_{CR}}{10^{-4}Mpc^{-3}}\right)^{-1}\ \
{\rm erg\  s^{-1}},
\label{LcR}
\end{equation}    
with $n_{CR}\sim$ a few times $10^{-4}$Mpc$^{-3}$ being the space density of 
objects contributing to the measured UHECR flux.  This constitutes less than
0.1\% of the maximum rotational power available.

\section{Curvature Losses and TeV Emission}
The dominant energy loss mechanism of the accelerating particles 
is curvature radiation \cite{Le00,BL00}.  In the limit of large suppression,
the maximum energy attainable can be expressed as \cite{Le00}:
\begin{equation}
\epsilon_{max}=3\times10^{19}\mu Z^{-1/4}M_9^{1/2}B_4^{1/4}
(\rho^2h/R_g^3)^{1/4}\ \ \ {\rm eV},
\label{Emx}
\end{equation}
where $\rho$ denotes the mean curvature radius of magnetic field lines, $\mu$ is the 
mass of accelerated particle in units of the proton mass,
and $Z$ is its charge.  The suppression factor is given in eq. (5) of 
ref. \cite{Le00}.  For the parameters corresponding to some of the candidate
UHECR sources (see ref. \cite{BG99} for a list),
the suppression factor for a proton lies in the range between 5 and 15, and 
the maximum acceleration energy given by equation (\ref{Emx}) is marginally sufficient
to accelerate protons to the required energies \cite{Le00,BL00}.  The 
constraints imposed on heavy nuclei are more relaxed (although heavy nuclei may be 
photo-disintegrated before escaping the system, owing to the interaction with 
the ambient radiation field present in those galaxies \cite{BL00}).

The emitted spectrum of curvature photons peaks in the TeV band \cite{Le00}.  
The average TeV flux per UHECR source can be estimated using equation (\ref{LcR}), 
and is shown \cite{Le00} to exceed the detection limit of present TeV experiments.
Given the IR luminosities and light profiles of the corresponding galaxies, as
well as upper limits on the luminosity of a continuum source, if present, it is
found that vacuum breakdown will not occur and that the TeV photons will 
escape the system.  Thus, prodigious
TeV emission appears to be a consequence of UHECR production by dormant AGNs.
As pointed out in ref. \cite{Le00}, low luminosity AGNs for which the 
breakdown criterion is not satisfied, may also produce CRs and curvature
emission.  However, the energy of the curvature photons will be degraded
to the sub-TeV range as a result of collisions with the ambient photons
produced in the accretion disk.  These sources should be potential targets for 
GLAST and MAGIC.

\section{Application to the Sgr A* Source}

Dynamical measurements indicate the presence of a black hole of mass 
$\sim 3\times10^6$ $M_{\sun}$ in the Galactic center \cite{EG97}.  Estimates of
the accretion rate (measured henceforth in Eddington units) based 
on observations of stellar winds in the vicinity of Sgr A* yield 
$\dot{m}\sim 10^{-3}$ and, assuming a radiative efficiency of 10 \%, 
accretion luminosity of $\sim 10^{40}$ erg s$^{-1}$ - several orders of 
magnitude greater than the observed value \cite{MDR}.  The low radiative output has 
been successfully explained in terms of an ADAF model \cite{NMG98}.  A 
viable model fit to the observed spectrum yields
values of $M$ and $\dot{m}$ consistent with those quoted above, and 
magnetic field pressure near equipartition, which, we estimate, 
corresponds to magnetic induction of the order of 10$^4$ G ($B_4\sim1$). 
Taking $a=M$, $h=R_g$ in eq. (\ref{V}), the maximum energy attainable by a nucleon
of charge $Z$ is $\epsilon_{max}\simeq 1.5\times10^{18} ZB_4$ eV; (since, as
shown immediately below, curvature losses are small or at most mild eq. [\ref{Emx}] is 
inapplicable in this case).  The curvature loss rate per
nuclei is $P\sim3\times10^5Z^6 B_4^4\mu^{-4}$ erg s$^{-1}$, and the corresponding 
radiative efficiency is $(P/c)(eZ\Delta V/h)^{-1}\sim0.5 Z^5B_4^3\mu^{-4}$.  
Thus, for $B_4<1$ curvature losses are not severe.

The maximum rotational power that can be extracted from the Galactic black hole 
is approximately $10^{41}B_4^2$ erg s$^{-1}$.  Denoting by $\eta_{CR}$ the CR production
efficiency, that is, the fraction of maximum power released as cosmic rays, we
obtain a flux of curvature photons at Earth (adopting a distance of 8.5 kpc to the 
Galactic center) of ${\cal F}_{\gamma}\sim 2\times10^{-5}\eta_{CR}B_4^5Z^5\mu^{-4}$ 
erg cm$^{-2}$ s$^{-1}$.  The curvature spectrum will peak at an energy 
$\epsilon_{\gamma}\sim200Z^3\mu^{-3}B_4^3$ GeV.  Adopting for illustration $B_4=1$
and $\eta_{CR}=10^{-3}$, roughly the efficiency inferred for the dormant AGNs (see above),
we obtain a CR power of about $10^{38}$ erg s$^{-1}$ for the Sgr A* source, with
a comparable $\gamma$-ray luminosity (a flux at Earth of $10^{-8}$ in cgs units), 
and peak energy of the $\gamma$-ray spectrum in the range 20 - 200 GeV, depending 
on the composition of accelerated particles.  This is well above detection limit of 
next generation $\gamma$-ray telescopes. We stress
that these results are highly uncertain in view of the strong dependence on 
the magnetic field strength.

Finally, we note that there have been claims \cite{AGASA,Cl00} that the measured CR flux 
near $10^{18}$ eV is anisotropic, indicating a strong CR source in the direction
of the Galactic center.  Whether this anisotropy can be accounted for by the model
discussed above remains to be checked; predicting the CR flux at Earth and the 
anisotropy amplitude is complicated by virtue of the diffusive nature of cosmic ray 
propagation in the Galaxy.  Detailed numerical simulations \cite{Cl00} appear 
to suggest that a CR source located in the vicinity of the Galactic center 
can account for the claimed anisotropy, provided the CR spectrum extends up to 
(or even slightly above) 10$^{18} Z$ eV.  This is compatible with the maximum
energy gain estimated above for the parameters corresponding to the ADAF model.

\section{acknowledgment}
I thank E. Boldt for discussions, and F. Aharonian for useful comments.


\begin{references}
\bibitem{BS99} Bhattacharjee, P.,\& Sigl, G. {\it Phys. Rep.} {\bf327} 109-247 (2000)
\bibitem{G} Greisen, K. {\it Phys. Rev. Lett.}, {\bf 16}, 748-751 (1966); 
\bibitem{ZK} Zatsepin, G.T., \& Kuzmin, V.A. {\it JETP Lett.} {\bf 4}, 78 (1966)
\bibitem{Ol00} Olinto, A.V. {\it Phys. Rep.} {\bf333-334}, 329-348 (2000)
\bibitem{BG99} Boldt, E. \& Ghosh, P., {\it Mon. Not. R. Astron. Soc.} 
{\bf 307}, 491-494 (1999)
\bibitem{M98} Magorian J., {\it et al.}, {\it Astron. J.} {\bf 115}, 2285-2305 (1998)
\bibitem{BZ77} Blandford, R.D. \& Znajek, R.L. {\it Mon. Not. R. Astron. Soc.}
{\bf 179}, 433-456 (1977)
\bibitem{TPM} Thorne, K.S., Price, R.M. \& MacDonald, D.A.
{\it Black holes: the membrane paradigm}, Yale Univ. Press, New Haven CN (1986) 
\bibitem{Le00} Levinson, A. {\it Phys. Rev. Lett.} {\bf 85}, 912-915 (2000)
\bibitem{BL00} Boldt E. \& Loewenstein M. {\it Mon. Not. R. Astron. Soc.} {\bf 316}, L29-L33 (2000)
\bibitem{EG97} Eckart, A., \& Genzel, R. {\it Mon. Not. R. Astron. Soc.} {\bf 284}, 576-598 (1997)
\bibitem{MDR} Mezger, P.G., Duschl, W.J., \& Zylka, R. {\it Astron. Astrophys. Rev.} {\bf 7} 289-388 (1996)
\bibitem{NMG98} Narayan, R., Mahadevan, R., \& Grindlay, L., {\it Astrophys. J.} {\bf 492}, 554-568 (1998)
\bibitem{AGASA} Hayashida, N., et al. {\it Astropart. Phys.} {\bf 10}, 303-311 (1999)
\bibitem{Cl00} Clay, R.W., Dawson, B.R., Bowen, J., \& Debes, M. {\it Astropart. Phys.}
{\bf 12}, 249-254 (2000)
\end{references}
\end{document}